\documentclass[prl,twocolumn,epsf,psfig]{revtex4}
\usepackage{graphicx}
\DeclareGraphicsExtensions{.pdf,.png,.gif,.jpg}
\usepackage{float}
\usepackage{subfig}
\usepackage{epsfig}
\usepackage{wrapfig}
\usepackage{bbold}

\begin{document}

\title{Bio-energy Transport as a Phonon Dressed Vibrational Exciton in Protein Molecules}

\author{Theja N. De Silva$^{1,2}$ and Peighton Bolt$^{1}$}
\affiliation{1. Department of Chemistry and Physics,
Augusta University, Augusta, Georgia 30912, USA;\\
2. Kavli Institute for Theoretical Physics, University of California, Santa Barbara, California 93106, USA.}

\begin{abstract}
Following the ideas of Davydov's soliton theory, we study the bio-energy transport in protein molecules. By using a quantum Brownian motion model for a phonon dressed vibrational exciton, we calculate the time-dependence on the mean square distance, diffusion coefficient, and energy of the vibrational exciton. We find the time-dependence by solving the quantum Langevin equation and find oscillatory behaviors due to the super-diffusive non-ohmic dissipation. We find that the vibrational exciton gains an overall energy due to the coupling to the phonon bath; it also dissipates its energy to the environment as it propagates. The amount of energy gain and the oscillatory features depend on both temperature and the phonon-vibron coupling.
\end{abstract}

\maketitle

\section{I. Introduction}

The energy needed for various biological phenomena and biochemical processes are captured during the hydrolysis of adenosine triphosphate (ATP)~\cite{atp1, atp2, atp3}. It is believed that about 0.43 ev of energy is released in these vital chemical events within living systems. This energy release is associated with the dissociation of phosphate-oxygen bonds in ATP~\cite{atpe1, atpe2}. The underlying mechanism of this energy release is not completely understood on a molecular basis. However, it is well understood that this bio-energy must transfer from one place to another in protein molecules to perform biological processes~\cite{bet1, bet2, bet3, bet4, bet5, bet6}. These biological processes include muscle contractions, DNA replications, neuro-electric pulse transfer, calcium pumps and sodium pumps all occur due to energy transfer through protein molecules. The theory of this energy transfer, now widely known as Davydov theory, was first proposed by Davydov~\cite{solm1, solm1A, solm1B}. It has been argued that the energy released from ATP is not sufficient to excite the electronic states of a molecule. The leading role in the energy transfer of proteins must be provided by the vibrations of atoms in $C=O$ bonds in the peptide group. However, the life time of such an isolated vibration (Amide-I vibration) is very short, thus Davydov suggested that the bio-energy is carried by a soliton mechanism in the protein molecule. In a quantum mechanical sense, a soliton is a wave packet that propagates in a medium while keeping its shape. The soliton mechanism suggests that the self-trapping Amide-I vibration, generated by the energy released in ATP hydrolysis, carries the energy through its interactions with acoustic phonons. These acoustic phonons are the harmonic lattice vibrations of the peptide molecular units in the protein.

There are many studies that support the energy transfer through soliton mechanism~\cite{solm2, solm3, solm4a, solm4b, solm5a, solm5b, solm5c, solm5d, solm5e, solm5f, solm5g, solm6a, solm6b, solm6c, solm7, solm8, solm9a, solm9b, solm10, solm11a, solm11b, solm11c}, however the Davydov soliton mechanism in $\alpha$-helix protein molecules needs some improvements to address many unanswered questions. It is well known that quantum mechanical effects, thermal effects, and disorder greatly effect the life time of solitons. A quantum mechanical perturbation calculation predicts that the life time of the Davydov soliton at biological temperature $300$K is too small to be responsible for biological processes~\cite{CS1, CS2}. Many improved models that involve soliton mechanism were proposed for the energy transfer~\cite{imm1a, imm1b, imm2a, imm2b, imm2c, imm3, imm4, imm5, imm6, imm7, new1, new2, new3}, but most of them were not able to explain the complete picture of underlying mechanism. Specifically, the soliton mechanism in the presence of classical phonons require strong exciton-phonon coupling for nonlinear dynamics~\cite{cst1, cst2, cst3}. Further, partially dressed vibrational excitons also require a strong exciton-phonon coupling for nonlinear soliton mechanics~\cite{pdp1, pdp2a, pdp2b, pdp3}. However, most \emph{ab initio} calculations estimate a smaller and negative coupling strength when compared with the values needed to satisfy nonlinear dynamics for classically treated phonons~\cite{abe1, abe2, abe3}. Although partially dressed vibrational excitons can be expected at low temperatures, fully dressed vibrational excitons are expected at biological temperatures. Therefore, for realistic systems, one must treat bio-energy transfer as a quantum mechanical propagation of fully dressed excitons at a weak coupling regime.

In a series of publications~\cite{cst2, vp1, vp2, vp3, vp4, vp5, vp6, vp7}, V. Pouthier \emph{et al.} have studied the bio-energy transport in a lattice of H-bonded peptide units using the framework of Davydov soliton mechanism. The idea is that the energy is transferred through phonon dressed vibrational excitons known as vibrons. The authors have studied the transport properties by calculating the reduced density matrix using the time-convolutionless master equation~\cite{vp6, vp7}. It is the purpose of this work to revisit the vibrational energy flow model and study the transport properties of peptide molecular chains by means of the quantum Langevin equation. Following Davydov's idea, we consider a simple model of a protein molecule where peptide groups are arranged in a periodic structure connected by hydrogen bonds. We treat quantum mechanical Brownian motion as a mechanism for the bio-energy transfer based on a self-trapped, narrow band, vibrational exciton in the presence of acoustic phonons generated by the vibration of peptide units. The motion of an effective vibrational exciton or Amide-I vibron is surrounded by elementary excitations of lattice vibrations or phonons. This vibron dressed with a virtual cloud of phonons is called a polaron. The concept of polaron is very common in studying mobility of impurities in materials. We investigate the physics of the fully phonon dressed vibron as an open system in the framework of quantum Brownian motion. In this context, the effective vibron plays the role of a Brownian particle moving in a thermal bath consisting of a collection of phonons satisfying Bose-Einstein statistics. We find that the transport properties are very sensitive to the temperature and the coupling between the vibron and the phonon bath. The calculated time-dependence on diffusion coefficient shows non-monotonic oscillatory behavior indicating that the vibron diffusion is not a simple diffusion nor coherent. Further, we find that the time-dependence of energy is also oscillatory. While dressed vibrons acquire an overall energy, they dissipate energy to the environment as it propagates. The overall gain in vibron energy depends on temperature and the coupling between the vibron and phonons.

The paper is organized into seven sections. Section II introduces a simple Frohlich type Hamiltonian as our model for the single vibron. Section III includes our derivation of the spectral density and the damping kernel through the self-correlation function. Section III also contains our arguement that the energy dissipation is super-ohmic and the vibron propagation carries memory effects. Section IV introduces the quantum Langevin equation and present its solution in section V. Section V also includes the derived time dependent transport quantities. Finally, section IV presents our results and we summarize our results with a discussion in section VII.

\section{II. The model}

The model describing the dynamics of vibrons is described by the Hamiltonian, $H = H_{ex} + H_{ph} + H_{int}$, where $H_{ex}$ represents the Boson-type vibrational excitation excited by the energy released in ATP, $H_{ph}$ represents the harmonic lattice vibrations of peptide molecular units, and $H_{int}$ represents the interaction between the vibron and the lattice vibrations. We consider a single vibron with an effective mass $m_{ex}$ and momentum $p_{ex}$,

\begin{eqnarray}
H_{ex} = \frac{p_{ex}^2}{2 m_{ex}}.
\end{eqnarray}

\noindent and a chain of peptide units connected by chemical bonds with an effective spring constant $k_s$, and mass $m$,

\begin{eqnarray}
H_{ph} = \sum_{n =1}^N \frac{m}{2} \frac{d^2x_n}{dt^2} + \sum_{n =1}^N \frac{1}{2}k_s (x_{n+1} - x_{n} -a)^2,
\end{eqnarray}

\noindent where $x_n$ is the position of $n$-th peptide unit in the chain and $a$ is the length of a peptide unit. Notice that we have introduced an effective mass to the vibron and consider the vibron as a particle moving along the peptide chain. This is done for the convenience in our theoretical formulation, however, we replace its effective mass at the end of our derivation using the initial kinetic energy of a single amide-I vibron $\epsilon_0 = \frac{\hbar^2}{2 m_{ex} l^2}$, where $l$ is the characteristic length scale in the problem. Assuming $|x_n(t) - \bar{x}_n| \leq a$ and changing the variables $x_n = \bar{x}_n + \phi_n$ with equilibrium position $\bar{x}_n = n a$, the lattice vibration part of the Hamiltonian can be written as

\begin{eqnarray}
H_{ph} = \sum_{n =1}^N \biggr[ \frac{p_n^2}{2 m} + \frac{k_s}{2} (\phi_{n+1} - \phi_n)^2 \biggr],
\end{eqnarray}

\noindent where $p_n = m (d\phi_n/dt)$. When the relative displacements of neighboring peptide units are small, $|\phi_{n+1} - \phi_n| \ll a$, we can take the continuum limit $\phi_n \rightarrow \phi(x)_{x = na}$ and the summation of discrete variables are then converted into an integral over the position $x$. Taking the Fourier transform of the displacement vector,

\begin{eqnarray}
\phi(x) = \frac{1}{L} \sum_k e^{ikx} \phi_k,
\end{eqnarray}

\noindent the phonon part of the Hamiltonian becomes,

\begin{eqnarray}
H_{ph} = \sum_{k} \biggr[ \frac{p_k p_{-k}}{2 \rho}  + \frac{1}{2} \rho \omega_k^2 \phi_k \phi_{-k} \biggr],
\end{eqnarray}

\noindent where $\omega_k = a k \sqrt{k_0/\rho}$ with $\rho = m/a$ and $k_0 = k_s/a$. Finally, the Hamiltonian can be represented in second quantized from in terms of bosonic operator $b_k^\dagger = \sqrt{\frac{m \omega_k}{2 \hbar}} (\phi_{-k} - \frac{i}{m \omega_k} p_k)$ as,

\begin{eqnarray}
H_{ph} = \sum_{k} E_k \biggr(b_k^\dagger b_k + \frac{1}{2}\biggr),
\end{eqnarray}

\noindent where $E_k = \hbar \omega_k$ and the usual bosonic operator $b_k^\dagger$ creates a phonon with wave vector $k$. The interaction part of the Hamiltonian $H_{int}$ now can be written in terms of bosonic operators,

\begin{eqnarray}
H_{int} = \frac{1}{L}\sum_{k} v_I(k) \rho_I (k) (b_k + b_{-k}^\dagger),
\end{eqnarray}

\noindent where $\rho_I (k) = \int_{-\infty}^{\infty} e^{ik\tilde{x}} \delta(\tilde{x} - x) d\tilde{x}$ is the density of vibron in the momentum domain and $v_I(k)$ is the interaction strength between an exciton at $x = \tilde{x}$ and the phonon gas. For the case of $kx \ll 1$, using $e^{ikx} \simeq 1 + ikx$, the final form of the system Hamiltonian can be simplified using the transformation $b_k \rightarrow b_k -v_I(k)/E_k$,

\begin{eqnarray}
H_{int} = H_{ex} + \sum_k E_k b_k^\dagger b_k + \sum_k \hbar g_k \pi_k x,
\end{eqnarray}

\noindent where we have defined effective coupling constant between exciton position and phonon momenta $g_k = k v_I(k)/\hbar$ and dimensionless momenta $\pi_k = i(b_k - b_k^\dagger)$. Notice that we have kept only the linear bosonic operators in the interaction term, thus our Hamiltonian is valid only for the weak coupling regime. In the polaronic physics literature, this weak coupling regime is known as Frohlich regime~\cite{fh1, fh2, fh3}. For the strong coupling regime, one needs to include the quadratic and higher order operators in the interaction term to generalize the model~\cite{gfh1, gfh2, gfh3, gfh4, gfh5}.  The Frohlich regime for the vibrons is justified by the \emph{ab initio} calculations for $\alpha$-helix fragments containing two H-bond peptide units~\cite{abe1, abe2, abe3}.

\section{III. Long time Super-diffusive non-ohmic dissipation}

The spectral density defined through the self-correlation function of the oscillators characterizes the dynamics of quantum Brownian motion of the dressed vibrons. The coupling between bath oscillator momenta $\pi_k$ and the dressed vibrons enters in the dynamics of the system via the self-correlation function $\emph{C}(\tau)$ of the oscillator environment~\cite{bp},

\begin{eqnarray}
\emph{C}(\tau) = \sum_k \hbar g_k^2 \langle \pi_k(\tau) \pi_k(0) \rangle,
\end{eqnarray}

\noindent where $\langle \pi_k(\tau) \pi_k(0) \rangle$ indicates the quantum or thermal average of the operator product $\pi_k(\tau) \pi_k(0)$  over the system Hamiltonian. Replacing oscillator momenta $\pi_k$ in terms of bosonic operators $b_k$ and using the phonon occupation number $n_k \equiv \langle b_k^\dagger b_k \rangle = 1/[e^{E_k/k_BT}-1]$, the self-correlation function can be cast as $\emph{C}(\tau) =  \nu(\tau) - i\lambda(\tau)$, where $k_BT$ is the dimensionless temperature with Boltzman constant $k_B$. The real part of the self-correlation function represents the noise,

\begin{eqnarray}
\nu(\tau) = \int_0^\infty J(\omega) \coth [\hbar \omega/2k_BT] \cos[\omega \tau] d\omega,
\end{eqnarray}

\noindent and the imaginary part of the self-correlation function represents the dissipation kernal,

\begin{eqnarray}
\lambda(\tau) = \int_0^\infty J(\omega) \sin[\omega \tau] d\omega,
\end{eqnarray}

\noindent where the spectral density $J(\omega)$

\begin{eqnarray}
J(\omega) = \sum_{k \neq 0} \hbar g_k^2 \delta(\omega - \omega_k).
\end{eqnarray}

\noindent Here $\omega_k = E_k/\hbar \equiv a \sqrt{k_0/\rho} k \equiv c k$. In order to derive an analytic expression for the spectral density, we assume interaction strength $v_I(k) \approx v_I(0)$ is momentum independent. As stated before, the weak coupling limit is justified by the \emph{ab initio} calculations for $\alpha$-helix fragments containing two H-bond peptide units~\cite{abe1, abe2, abe3}. Although theoretical estimates give small negative coupling constants, the sign of the interaction does not effect the dynamics as the spectral density depends on the square of the coupling constant. Further, in the presence of a large number of oscillators, the discrete sum can be converted into an continuous integral over the momenta, $\frac{1}{N}\sum_k \rightarrow \frac{a}{2 \pi}\int dk$. The resulting spectral density is,

\begin{eqnarray}
J(\omega) = \frac{v_I(0)^2 a}{2 \pi \hbar c^3} \omega^2.
\end{eqnarray}

\noindent Notice that the spectral density \emph{does not} depend linearly on the phonon frequency, thus the dissipation is not ohmic but super-ohmic. This super-ohmic dependence has been found in many other physical systems; such as an impurity immersed in a Luttinger liquid~\cite{sol1, sol2}, a bright soliton in a superfluid~\cite{sos}, and a Bose-impurity immersed in a Bose-Einstein condensate~\cite{bp, lampo}. This super-ohmic behavior is due to the linear momentum dependence of the phonon spectrum. The damping kernel, which is related to the life time of the dressed exciton can be written using spectral density,

\begin{eqnarray}
\Gamma(\tau) = \frac{1}{m_{ex}} \int_0^\infty \frac{J(\omega)}{\omega} \cos[\omega \tau] d\omega.
\end{eqnarray}

\noindent The integration can be completed by introducing a frequency cut-off $\Lambda$ as an upper bound,

\begin{eqnarray}
\Gamma(t) = \frac{v_I(0)^2 a}{2 \pi \hbar c^3 m_{ex}} \biggr(\frac{\Lambda}{t} \sin[\Lambda t] + \frac{1}{t^2} \cos[\Lambda t] - \frac{1}{t^2} \biggr).
\end{eqnarray}

\noindent The introduction of a cut-off frequency $\Lambda$ is reasonable as we are interested in the long-time dynamics of the dressed vibron. In addition, this natural cut-off frequency distinguishes the linear phonon spectrum from the quadratic spectrum and the low-frequency portion of the damping kernel is responsible for the long-time behavior. The most important feature of the damping kernel is that it is non-local in time. In other words, the dressed vibron at position $x$ at time $t$ depends on its past trajectory. This non-local time dependence indicates that the dynamics of a vibron always carries memory effects. In the following section, we investigate time dependent transport properties through a generalized quantum Langevin equation.

\section{IV. The generalized Quantum Langevin Equation}

Even though our Hamiltonian does not explicitly include the time-dependent forces, the energy dissipation occurs in non-linear systems through coupling between different degrees of freedoms. The statistical dynamics governed by the fluctuations are given by the generalized Langevin equation~\cite{gle1a, gle1b, gle2}. Starting with the Heisenberg equations of motions for the operators $\eta (t) = x(t), p(t), b_k(t), b^\dagger_k(t)$,

\begin{eqnarray}
\frac{d\eta}{dt} = \frac{i}{\hbar} [H, \eta(t)],
\end{eqnarray}

\noindent with $[H, \eta(t)]$ being the commutator between the system Hamiltonian and the time-dependent operator $\eta(t) = \eta \exp(-i Ht/\hbar)$, and combining these equations, one can derive the generalized Langevin equation for the system. The derivation is straight forward, however it is lengthy and tedious. The details can be found in ref. \cite{lampo}. The resulting generalized Langevin equation has the form,

\begin{eqnarray}
m_{ex} \frac{d^2x(t)}{dt^2} - \int_0^t \lambda (t-s) x(s) ds = \xi (t),
\end{eqnarray}

\noindent where $\xi(t) = \sum_k \hbar g_k \pi_k(t)$ is the generalized quantum Brownian force factor and $\lambda (t)$ is the dissipation kernel introduced in the previous section. The dissipation kernel is related to the damping kernel $\Gamma(t)$ according to the relateion $ m_{ex} (d\Gamma/dt) = -\lambda(t)$.

\section{V. Time Evolution of the Vibron's Mean Square Distance, Diffusion Coefficient, and Energy}

The time dependence on the dressed vibron energy $E(t) = \langle p^2(t) \rangle/(2m)$, mean square distance $X_{m}(t) = \langle x^2(t) \rangle$, and diffusion coefficient $D(t) = \frac{d}{dt}  \langle x^2(t) \rangle/2$ can be calculated through the solution $x(t)$ of generalized Langevin equation. Using Laplace transformation techniques and assuming an initial condition $x(t=0) = 0$, the solution to the Langevin equation can be written as,

\begin{eqnarray}
x(t) = G(t) \frac{dx(0)}{dt} + \frac{1}{m_{ex}} \int_0^\infty G(t-s) \xi(s) ds,
\end{eqnarray}

\noindent where the Green's function $G(t)$ is defined through the laplace transformation,

\begin{eqnarray}
\mathcal{L}_z\{G(t)\} = \frac{1}{z^2 + z \mathcal{L}_z\{\Gamma(t)\}},
\end{eqnarray}

\noindent and satisfy the boundary condition $G(t = 0) =0 $ and $\frac{dG}{dt}|_{t = 0} = 1$. By calculating $\langle |x(t)|^2 \rangle$ and $\langle |dx(t)/dt|^2 \rangle$ and using $ \langle \xi(s) \xi(\sigma) + \xi(\sigma) \xi(s) \rangle = 2 \hbar \nu(s-\sigma)$, we derive close form expressions for $E(t)$, $X_{m}(t)$, and $D(t)$ for the time dependence. By defining the natural frequency for peptide units $\omega_0 = \sqrt{k_s/m}$, with effective spring constants for chemical peptide bonds $k_s$ and mass $m$, we scale the energy by $\hbar \omega_0$, the length by $l = \sqrt{\frac{\hbar}{m \omega_0}}$, and the time by $\omega_0$ (i.e. $t \rightarrow t\omega_0$, $\Lambda \rightarrow \Lambda /\omega_0$, $T \rightarrow k_BT/\hbar \omega_0$ etc.). This allow us to construct the dimensionless equations for the energy, mean square distance, and the diffusion coefficient,

\begin{eqnarray}
\frac{E(t)}{\hbar \omega_0} = \frac{E(0)}{\hbar \omega_0} \biggr(\frac{dG(t)}{dt}\biggr)^2 + \gamma \delta(t-t^\prime) F(t, t^\prime),
\end{eqnarray}

\begin{eqnarray}
\frac{m_{ex}X_{m}(t)}{\hbar} = \gamma  \int_0^t ds \int_0^t d\sigma G(t-s)G(t-\sigma) I(s-\sigma) \\ \nonumber
+ \frac{E(0)}{\hbar \omega_0} G^2(t),
\end{eqnarray}

\noindent and

\begin{eqnarray}
\frac{m_{ex}D(t)}{\hbar} = \frac{m_{ex}}{\hbar} \frac{d}{dt}X_{m}(t) .
\end{eqnarray}

\noindent Here we have defined a dimensionless coupling constant $\gamma = \frac{\epsilon_0 v^2_I(0)}{2 \pi  (\hbar \omega_0)^3} \biggr( \frac{l}{a}\biggr)^2$, where the effective mass is replaced by the initial kinetic energy of the amide-I vibron $\epsilon_0 = \frac{\hbar^2}{2 m_{ex} l^2}$. The two dimensionless functions are defined as,

\begin{eqnarray}
F(t, t^\prime) = \frac{d}{dt} \frac{d}{d t^\prime} \int_0^t ds \int_0^{t^\prime} d\sigma G(t-s) G(t^\prime-\sigma) I(s-\sigma),
\end{eqnarray}

\noindent and

\begin{eqnarray}
I(t) = \int_0^{y_0} y^2  \coth[y/2T] \cos[yt] dy.
\end{eqnarray}

\noindent The cut-off frequency $y_0 = \Lambda/\omega_0$ appears in $\mathcal{L}_z\{G(t)\}$. The laplace transformation of the dimensionless Green's function $G(t)$ reads,

\begin{eqnarray}
\mathcal{L}_s\{G(t)\} = \frac{1}{s^2 +  y_0 \gamma s^2 \ln[1 + y_0^2/s^2]}.
\end{eqnarray}

\noindent Equations (20)-(22) complete the main theoretical formulation of the present study. All physical quantities, such as the time, energy, frequency, and temperature are now written as dimensionless quantities. The boundary condition of the Green's function $G(t)$ guarantees that $E(t)\rightarrow E(0)$, $X(t) \rightarrow 0$, and $D(t) \rightarrow 0$ in the limit $t \rightarrow 0$.

\section{VI. The Results}

In order to evaluate the time dependence on dimensionless energy, mean square distance, diffusion coefficient, three input parameters; the dimensionless cut-off frequency $y_0$, the effective dimensionless coupling parameter $\gamma$, and the dimensionless initial energy $E(0)$ are needed. Using widely accepted values for the $\alpha$-helix protein molecules~\cite{vp6, params}, we have $E(0) = 3.36 \times 10^{-20} J$ and $\hbar \omega_0 \sim (1 - 13) \times 10^{-22} J$. The room temperature in dimensionless form $k_BT/\hbar \omega_0$ is about $4$. For a lattice of H-bonded peptide units, the vibron bandwidth is smaller than the phonon cut-off frequency $2 \omega_0$~\cite{vp6}. Therefore, we set our cut-off frequency $\Lambda = 2 \omega_0$ for the longer time dynamics. Depending on the dimensionless coupling constant $\gamma$, we can define three interaction regimes as the, weak coupling limit ($\gamma \ll 1$), moderate coupling limit ($\gamma \sim 1$), and strong coupling limit ($\gamma \gg 1$).

The Green's function defined through the Eq. (25) does not have an analytical form. As we have introduced a dimensionless frequency cut-off $y_0$ in our formalism, our theory is valid only for longer-time dynamics. At the zero interaction limit where the coupling constant $\gamma = 0$, the green's function has a simple linear form $G (t) = t$. Notice that the non-interacting Green's function is independent of the cut-off and it obeys the boundary conditions. In the presence of interaction, we numerically calculate the longer time Green's function. We find that the Green's function has an oscillatory behavior at shorter-times and a linear behavior at long-times. As we are interested in the longer time dynamics, we present our results in the longer time scale where the Green's function is in linear time dependence region. Notice that the Green's function is independent of the temperature, but strongly depends on the coupling constant.

\begin{figure}
\includegraphics[width=\columnwidth]{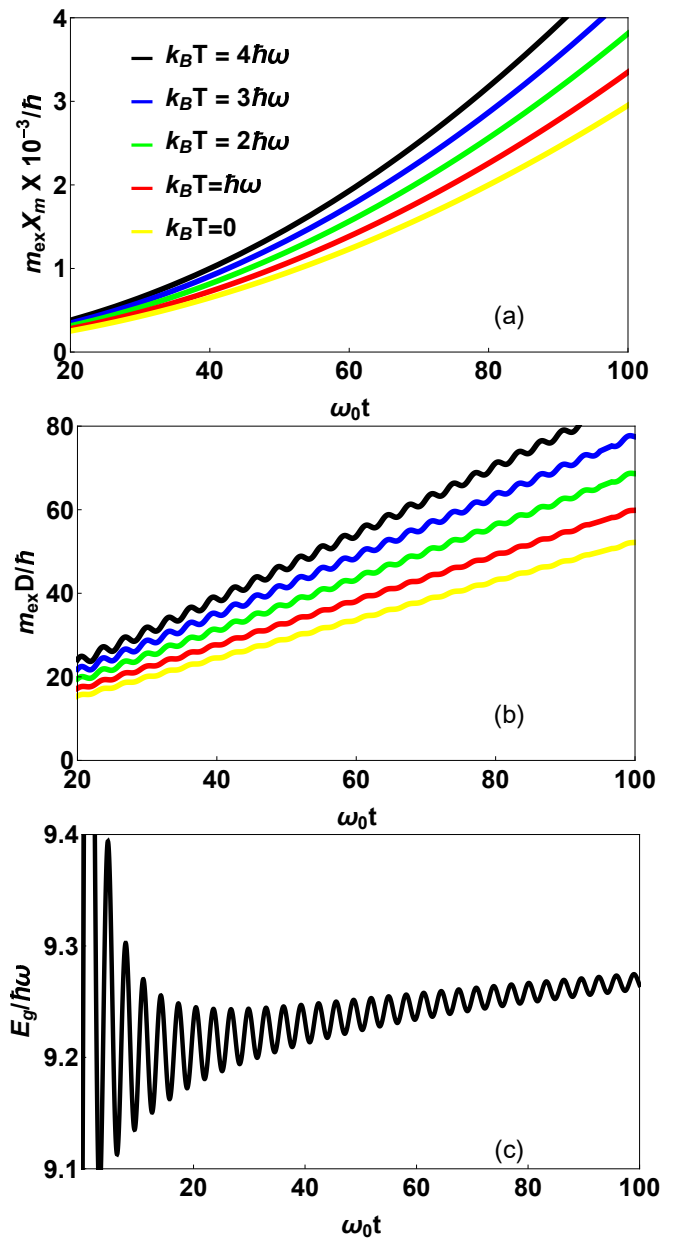}
\caption{(color online) The time dependence of the dimensionless mean square distance ($X_{m})$ in panel (a), diffusion coefficient ($D$) in panel (b), and energy ($E$) in panel (c), at different temperatures. The coupling parameter has been fixed to be $\gamma = 0.5$. The results are shown from zero temperature to room temperature and the color code is given in panel (a).}\label{TvarF}
\end{figure}

\begin{figure}
\includegraphics[width=\columnwidth]{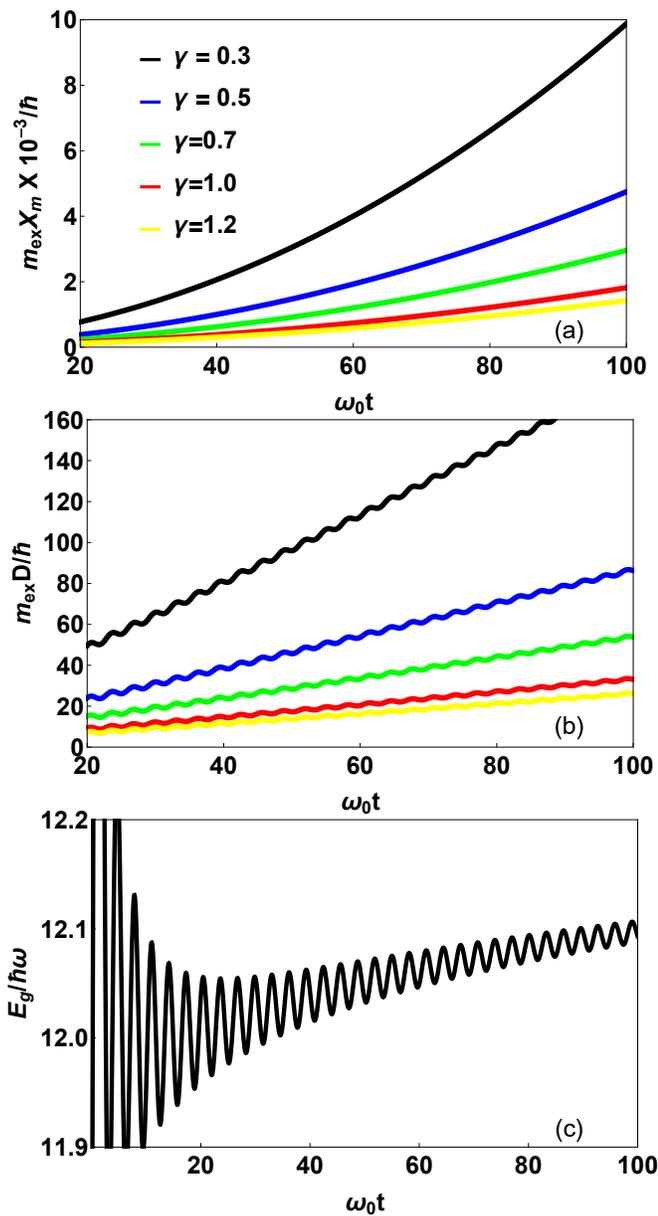}
\caption{(color online) The time dependence of the dimensionless mean square distance ($X_{m}$) in panel (a), diffusion coefficient ($D$) in panel (b), and energy ($E$) in panel (c) at different coupling strengths. The temperature has been fixed to be about room temperature $k_BT = 4 \hbar \omega_0$. The color code is given in in panel (a).}\label{IvarF}
\end{figure}

The FIG.~\ref{TvarF} and FIG.~\ref{IvarF} show our results at a longer time regime for a set of representative values of input parameters. While FIG.~\ref{TvarF} shows the temperature dependence, the FIG.~\ref{IvarF} shows the coupling dependence of transport properties as a function of dimensionless time. As seen from the graphs, both temperature and coupling dependence show similar time dependent behavior.

\noindent The mean square distance $X_m$ monotonically increases as one increases the time or the temperature, however, it monotonically decreases with the coupling parameter. The behavior of $X_m$ indicates the power law dependence on time and the power law exponent increases with increasing the temperature or decreasing the coupling parameter. This indicates that the dressed vibrons move at a faster rate at higher temperatures and smaller couplings with phonons. The non-linear time dependence of the mean square distance is a signature of the super-diffusion caused by the memory effects discussed before~\cite{bookME}.

Notice that the diffusion coefficients shown in panel (b) of FIG.~\ref{TvarF} and FIG.~\ref{IvarF} are neither linear with respect to time nor time-independent. Essentially, the energy dissipation is not a simple coherent nor an incoherent normal diffusive transfer. The diffusion coefficient shows an oscillatory behavior with time and the oscillation amplitude is larger at larger temperatures and at smaller coupling parameters. The time-average diffusion coefficient is linear indicating somewhat coherent energy transfer resulting from a wavelike motion of the vibron. This oscillatory behavior of the diffusion coefficient contrasts with the linear time-dependence predicted in previous generalized master equation approach~\cite{vp6}.

In Panel (c) of FIG.~\ref{TvarF} and FIG.~\ref{IvarF}, we present the time dependence of vibron energy for a set of two representative values of the coupling parameter at room temperature. Here the time-dependent energy $E_g$ is defined as $E_g = E(t) - E(0) (dG/dt)^2$. In fact, the second term here is time-independent in the longer time scale as the Green's function $G(t)$ is linear. The time dependence of the energy also shows an oscillatory behavior. Initially, the vibron gains energy from the environment and then oscillates rapidly before it approaches to an asymptotic limit at a large time. The gain in energy is due to the flow of energy from the phonon bath to the vibron. This energy back flow is direct evidence of memory effects~\cite{ebf}. The rate of energy dissipation/gain is larger at higher temperatures and at smaller coupling parameters. Even though the energy dissipation/gain is higher, the dressed vibron  moves a longer distance for higher temperatures and smaller coupling parameters, as evidenced by the mean square distance values in panel (a) of FIG.~ \ref{TvarF} and FIG.~\ref{IvarF}. The non-monotonic, oscillatory time-dependence of the energy indicates that the dressed vibron does not just acquires its energy, but also dissipates energy to the environment as it propagates.

\section{VII. Conclusions}

Inspired by the Davydov's soliton theory, we treated a vibrational exciton as a bio-energy carrier and discussed the physics of energy transport in protein molecules. We used a quantum Brownian motion model for phonon dressed vibrational exciton and calculated the time-dependence on dynamical properties using quantum Langevin equation. We found that the Brownian motion of the vibrational exciton in the presence of vibron-phonon interaction yields a super-diffusive transport behaviour to the vibron. This super-diffusive behavior gives an oscillatory time dependence for both the diffusion coefficient and energy. However, the time average diffusion coefficient is linear indicating that the time average energy transport is coherent. We find that the vibrational exciton gains an overall energy, however, it also dissipates its energy to the environment as it propagates. We find that the transport properties are very sensitive to temperature and the phonon-vibron coupling.

In this work, we neglected the phonon-phonon interactions, anharmonic effects, and lattice effects. We do not expect such effects to change the qualitative features of the transport properties. While phonon-phonon interaction renormalizes the phonon frequency, anharmonic effects may weaken the oscillatory behavior. Further, we have restricted ourselves to a single exciton dressed in a bath of phonons and investigated the effect of its coupling to the phonons and quantum fluctuations on the dynamics. Studying the interacting many excitons in a phonon bath is a challenging task that requires powerful theoretical tools to unravel the dynamics. The restriction to a single exciton is reasonable for energy transfer in a biological systems and makes the problem tractable to a mostly analytical treatment, allowing us to gain important physical insight.

\section{VIII. Acknowledgments}

The authors acknowledges the support of Augusta University and the hospitality of KITP at UC-Santa Barbara. A part of this research was completed at KITP and was supported in part by the National Science Foundation under Grant No. NSF PHY11-25915.

\end{document}